\documentclass[12pt]{iopart}
\usepackage{graphicx}
\begin{document}

\title[Open Charm Production at STAR]{Open Charm Production at STAR}

\author{Haibin Zhang\dag\ (for the STAR Collaboration)
\footnote[3]{For the full author list and acknowledgements, see
Appendix ``Collaboration" in this volume.} }

\address{\dag\ Physics Department, Brookhaven National Laboratory, Upton, NY, 11953, USA\\
E-mail: haibin@bnl.gov}

\begin{abstract}
We present the open charm spectra at mid-rapidity from direct
reconstruction of $D^0$, $D^*$ and $D^\pm$ in d+Au collisions at
$\sqrt{s_{NN}}$=200 GeV using the STAR detector at RHIC. The
indirect electron/positron measurements via charm semileptonic
decays in p+p and d+Au collisions are also reported. The total
$c\bar{c}$ cross section per nucleon-nucleon collision is
extracted from both direct and indirect measurements and are
consistent with each other. By combining the $D^0$ and
semileptonic measurements together, the cross section of
1.4$\pm$0.2$\pm$0.4 mb is higher than expectations from PYTHIA and
other pQCD calculations. The open charm $p_T$ distribution from
direct measurements covers the $p_T$ range up to $\sim$10 GeV/c
and follows a power-law distribution.

\end{abstract}


\submitto{\JPG}


\section{Introduction}
Due to the relatively large mass of the heavy quarks which
requires large energies for their creaction, the study of heavy
flavor productions provides a unique tool to test the perturbative
Quantum Chromodynamics (pQCD) calculations for heavy quark
predictions~\cite{frixione}. The energy dependence of the total
$c\bar{c}$ cross section measured from experiments is an important
variable to constrain various parameters in the pQCD
calculations~\cite{vogt}. It is predicted that the charm quarks
are mostly produced from the initial fusion of gluons in a high
energy hadron-hadron collision~\cite{lin}. Thus, the measurement
of open charm hadron production is an important tool to study the
initial parton distribution function in the colliding nucleons. In
relativistic heavy ion collisions, the open charm $p_T$ spectrum
may be modified due to parton energy loss in the deconfined hot
and dense medium. However, this energy loss is predicted to be
significantly smaller compared to light quark hadrons due to the
``deadcone effect"~\cite{dima}. As a consequence, an increased
$D/\pi$ ratio as a function of $p_T$ is expected in central A+A
collisions compared to p+p collisions at the same beam energy.
Therefore, the measurement of the open charm $p_T$ spectrum can
provide insights into the properties of the matter formed in the
relativistic heavy ion collisions. Furthermore, the production of
open charm hadrons in the same collision system provides a
comparison baseline to test the $J/\psi$ production mechanism in
the relativistic heavy ion collisions at RHIC energies since some
theories predict a $J/\psi$ enhancement due to the charm quark
coalescence effect~\cite{peter1,peter2,j1,j2,j3} while others
predict a $J/\psi$ suppression due to the color screening
effect~\cite{j4}.

\section{Measurements}
The data used for the direct $D^{0}$, $D^{*}$ and $D^{\pm}$
reconstruction and the charm semileptonic decay analysis were
taken during the 2003 RHIC run in d+Au and p+p collisions at
$\sqrt{s_{NN}}$=200 GeV in the STAR experiment. A minimum bias
d+Au collision trigger was defined by requiring at least one
spectator neutron in the Au beam outgoing direction depositing
energy in the Zero Degree Calorimeter (ZDC). A total of 15 million
minimum bias triggered d+Au collision events were used in the
$D^{0}$, $D^{*}$ and $D^{\pm}$ analysis. The data samples used in
the electron analysis in d+Au and p+p collisions were described in
Ref.~\cite{data}.

\subsection{Direct $D$ Meson Reconstruction}
The primary tracking device of the STAR detector is the Time
Projection Chamber (TPC) which measures the kinematics of charged
particles. Via the energy loss ($dE/dx$) in the TPC, charged kaons
and pions can be identified with their momenta up to $\sim$0.75
GeV/c. Due to the relatively small decay lengths
($c\tau\sim$100$\mu$m) of the $D^{0}$, $D^*$ and $D^{\pm}$, the
limited TPC track projection resolution does not allow the decay
topology identification by displaced vertices for open charm
mesons from the primary collision vertex. Thus the event-mixing
technique was used to reconstruct the invariant mass spectra of
the $D$ mesons. Detailed explanation of this method was addressed
in Ref.~\cite{mix,mix2,mix3}.

\begin{figure}[htp]
\centering
\includegraphics[height=23pc,width=30pc]{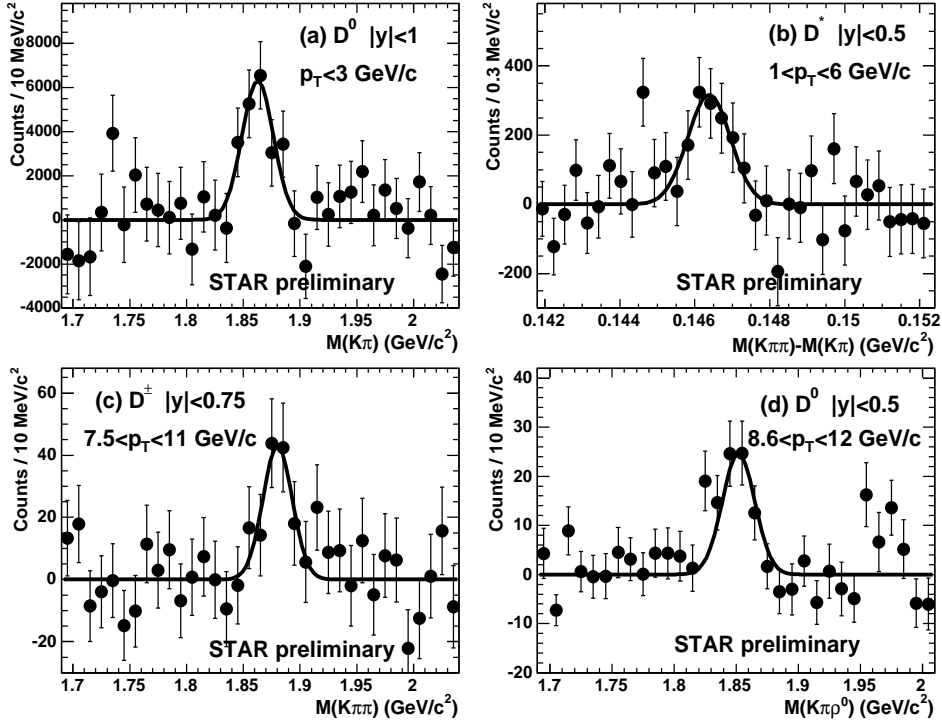}
\caption{The invariant mass distribution after the mixed-event
background subtraction and a further linear background subtraction
of $M(K\pi)$ for the low $p_T$ ($p_T<$3 GeV/c) $D^0+\bar{D^0}$ in
Panel (a), $\Delta M=M(K\pi\pi_s)-M(K\pi)$ for the $D^{*\pm}$ in
Panel (b), $M(K\pi\pi)$ for the $D^\pm$ in Panel (c),
$M(K\pi\rho^0)$ for the high $p_T$ (8.6$<p_T<$12 GeV/c)
$D^0+\bar{D^0}$ in Panel (d), respectively.}\label{fig:charmpeaks}
\end{figure}

The low $p_T$ ($p_T<$3 GeV/c) $D^0$ meson was reconstructed
through the decay of $D^0\rightarrow K^- \pi^+$
($\bar{D^0}\rightarrow K^+ \pi^-$) which has a branching ratio of
3.83\%. The oppositely charged $K\pi$ invariant mass distribution
after the combinatorial background subtraction from event-mixing
and a linear residual background subtraction is shown in Panel (a)
of Fig.~\ref{fig:charmpeaks} with a clear $D^0$ signal visible.
This distribution was fit with a Gaussian function and the mass
and width of the $D^0$ was found to be 1863$\pm$3 MeV/c$^2$
(1864.5$\pm$0.5 MeV/c$^2$ in the Particle Data Group~\cite{pdg})
and 13.8$\pm$2.8 MeV/c$^2$, respectively.

The $D^{*\pm}$ mesons were reconstructed through the decay of
$D^{*+}\rightarrow D^0 \pi^+_s$ ($D^{*-}\rightarrow \bar{D^0}
\pi^-_s$) with a branching ratio of 68\%. Specific treatment for
the soft pion daughter $\pi^\pm_s$ was performed which required
its $p_T$ and $p$ between 0.1 to 1 GeV/c and the ratio of the
$D^0$ to $\pi_s$ momentum $p(D^0)/p(\pi_s)>$9 since the $\pi_s$
has an average momentum of about 50 MeV/c. The invariant mass of
the kaon and pion candidates $M(K\pi)$ was calculated and required
to satisfy the $D^0$ mass 1.82$<M(K\pi)<$1.9 GeV/c$^2$. Then a
soft pion candidate with its charge opposite to that of the kaon
candidate was combined with the $D^0$ candidate to calculate the
invariant mass of a $D^*$ candidate $M(K\pi\pi_s)$. Panel (b) of
Fig.~\ref{fig:charmpeaks} shows the distribution of the invariant
mass difference $\Delta M=M(K\pi\pi_s)-M(K\pi)$ after background
subtractions. A clear signal is seen around the nominal value of
$M(D^*)-M(D^0)$. This distribution was fit with a Gaussian
function and the mass difference and the width was found to be
146.37$\pm$0.15 MeV/c$^2$ (145.42$\pm$0.05 MeV/c$^2$ in the PDG)
and 0.57$\pm$0.16 MeV/c$^2$, respectively.

The $D^\pm$ signal was reconstructed through the decay of
$D^+\rightarrow K^-\pi^+\pi^+$ ($D^-\rightarrow K^+\pi^-\pi^-$)
with a branching ratio of 9.1\%. The invariant mass distribution
of $M(K\pi\pi)$ after background subtractions is shown in Panel
(c) of Fig.~\ref{fig:charmpeaks} with the $D^\pm$ rapidity
$|y|<$0.75 and 7.5$<p_T<$11 GeV/c. A Gaussian function was fit to
this distribution and the mass and width was found to be
1880$\pm$5 MeV/c$^2$ (1869.3$\pm$0.5 MeV/c$^2$ in the PDG) and
16.2$\pm$8 MeV/c$^2$, respectively.

The high $p_T$ (8.6$<p_T<$12 GeV/c) $D^0$ was independently
reconstructed through the decay of $D^0\rightarrow K^-\pi^+\rho^0$
($\bar{D^0}\rightarrow K^+\pi^-\rho^0$) which has a branching
ratio 6.2\%. The analysis was similar to that for the low $p_T$
$D^0$ except that an additional $\pi^+\pi^-$ pair with its
invariant mass 0.62$<M(\pi^+\pi^-)<$0.86 was combined with the
selected kaon and pion candidates for a $D^0$ candidate. The
invariant mass distribution for $M(K\pi\rho^0)$ (=$M(K\pi\pi\pi)$)
after background subtractions is shown in Panel (d) of
Fig.~\ref{fig:charmpeaks}. This distribution was fit with a
Gaussian function and the mass and the width was found to be
1850$\pm$5 MeV/c$^2$ and 13.6$\pm$4 MeV/c$^2$, respectively.

\subsection{Single Electron Analysis}
\begin{figure}[htp]
\centering
\includegraphics[height=21pc,width=30pc]{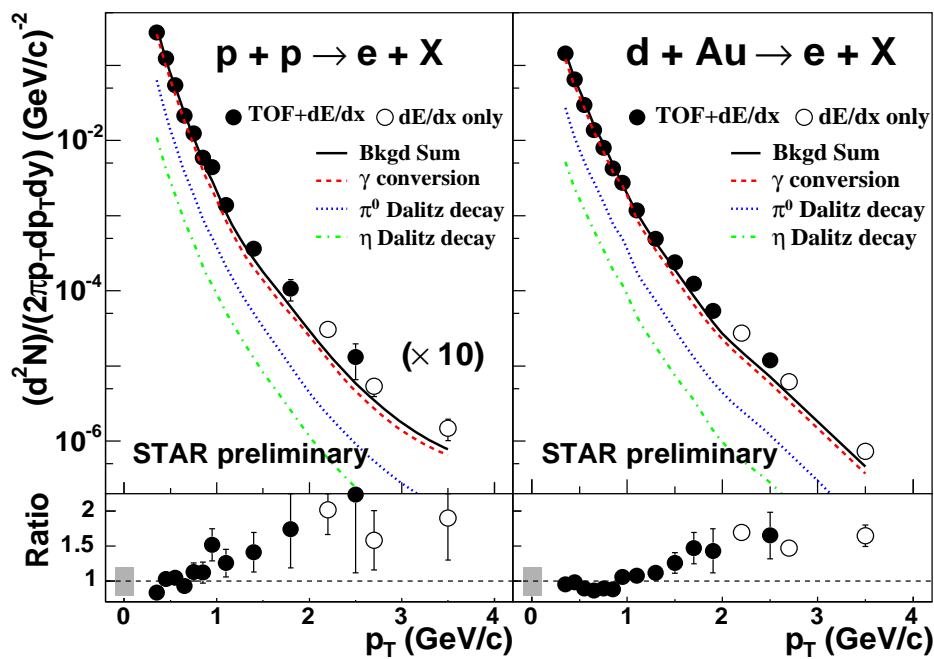}
\caption{Upper panels: Electron distributions from p+p (left) and
d+Au (right) collisions. Bottom panels: Ratios of inclusive
electrons over the total backgrounds. The gray bands represent the
systematic uncertainties.}\label{fig:electron}
\end{figure}

A prototype time-of-flight system (TOFr)~\cite{tof1,tof2} based on
the multi-gap resistive plate chamber technology was installed in
STAR with an azimuthal angle coverage $\Delta \phi\simeq\pi/30$
and pseudorapidity range -1$<\eta<$0. Besides its capability of
hadron identification, electrons/positrons could be identified at
$p_T<$3 GeV/c by the combination of velocity ($\beta$) from TOFr
and $dE/dx$ from TPC measurements. In addition, electrons can also
be identified with 2$<p_T<$4 GeV/c in TPC since hadrons have lower
$dE/dx$ due to the relativistic rise of the $dE/dx$ of electrons.
The inclusive spectra for the low $p_T$ electrons/positrons
measured from TOFr+TPC are shown in Fig.~\ref{fig:electron} as
solid symbols and the spectra for higher $p_T$ electrons from TPC
only are shown as open circles for p+p (left) and d+Au (right),
respectively.

The dominant sources of the electron background are the photon
conversions $\gamma\rightarrow e^+e^-$ (dashed curves in
Fig.~\ref{fig:electron}) and $\pi^0\rightarrow\gamma e^+e^-$
Dalitz decay (dotted curves) and $\eta$ Dalitz decay (dash-dotted
curves). To measure these photonic electron background spectra,
the invariant mass of the $e^+e^-$ pairs were constructed from an
electron (positron) in TOFr and every positron (electron)
candidate in TPC. The sum of these photonic background is shown as
the solid curves in Fig.~\ref{fig:electron}. In the bottom panel
of Fig.~\ref{fig:electron}, the ratio of the inclusive
electron/positron spectra and the background is shown and clear
signal excesses are visible with $p_T>$1 GeV/c.

\section{Results}

\subsection{Total $c\bar{c}$ Cross Section}
From the direct low $p_T$ $D^0$ reconstruction in the d+Au
collisions, the invariant yield $d^2N/2\pi p_Tdp_Tdy$ as a
function of $p_T$ after efficiency and acceptance correction was
extracted in four $p_T$ bins at $p_T<$3 GeV/c. Using an
exponential fit to the invariant yield in transverse mass ($m_T$),
the midrapidity yield $dN/dy$ for $D^0$ was found to be
0.028$\pm$0.004($stat.$)$\pm$0.008($sys.$). We also performed a
fit with the combined results of $D^0$ and electron distributions
in d+Au collisions, assuming that the $D^0$ spectrum follows a
power law in $p_T$ and that the remaining electrons after the
background subtractions are charm semileptonic decays. The yield
difference between the above two fitting methods is much smaller
than the statistical uncertainties.

\begin{figure}[htp]
\centering
\includegraphics[height=21pc,width=30pc]{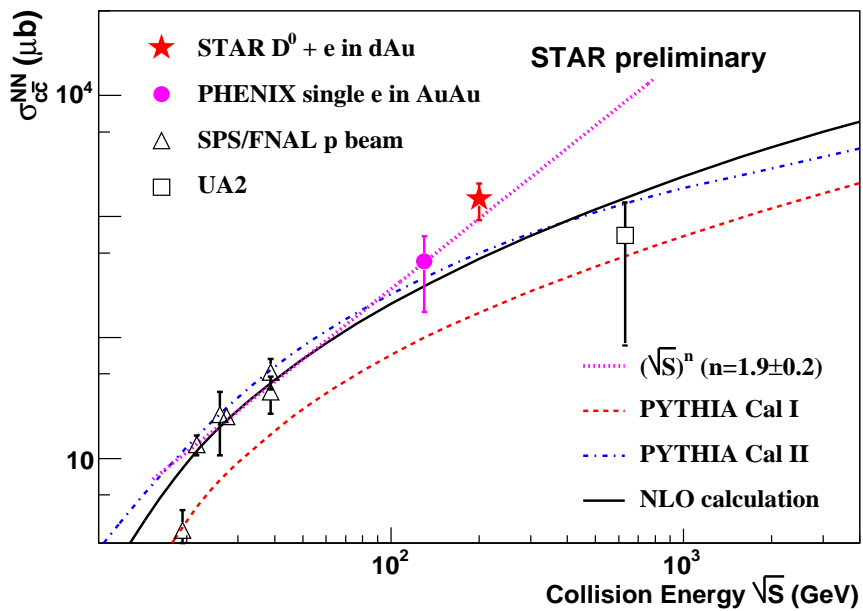}
\caption{The total $c\bar{c}$ cross section per nucleon-nucleon
collision vs. the collision energy. The dotted line is a power-law
fit. The dashed and dash-dotted curves are PYTHIA calculations
with different options. The solid curve is a NLO pQCD
calculation.}\label{fig:crosssection}
\end{figure}

We used the ratio $R=N(D^0)/N(c\bar{c})$=0.54$\pm$0.05 from
$e^+e^-$ collisions~\cite{pdg} to convert the $D^0$ yield to total
$c\bar{c}$ yield. The d+Au number of binary collisions $N_{bin}$
and the p+p inelastic scattering cross section was used to convert
the $dN^{c\bar{c}}/dy$ in d+Au collisions in to
$d\sigma^{c\bar{c}}/dy$ in p+p collisions. A factor of
4.7$\pm$0.7~\cite{vogt2,sjostrand} was used to convert the
$d\sigma/dy$ at midrapidity to the total cross section. The total
charm cross section per nucleon-nucleon interaction for d+Au
collisions at 200 GeV is 1.3$\pm$0.2$\pm$0.4 mb from $D^0$ alone
and 1.4$\pm$0.2$\pm$0.4 mb from the combined fit of $D^0$ and
electrons.

The beam energy dependence of the charm cross section from this
analysis is depicted in Fig.~\ref{fig:crosssection} by the star
symbol and compared to PHENIX~\cite{phenix}, UA2~\cite{ua2}, FNAL
and SPS~\cite{sps} measurements. The dotted line is a power law
fit, $\sigma^{NN}_{c\bar{c}}\propto(\sqrt{s})^n$, to the data
points with $n$=1.9$\pm$0.2, while $n\sim$0.3(0.5) had been
observed for charged hadron (pion) productions~\cite{j1}. This
indicates a harder behavior of the underlying process going from
charged hadron production to charm dominated processes. The dashed
and dash-dotted curves depict PYTHIA calculations with and without
higher order processes~\cite{sjostrand}. The solid curve depicts
the next-to-leading order (NLO) pQCD calculation from Ref.
~\cite{vogt2}. At $\sqrt{s}$=200 GeV, both calculations
underpredict the total charm cross section by at least a factor of
3.

\subsection{Open Charm $p_T$ Spectrum}
\begin{figure}[htp]
\centering
\includegraphics[height=21pc,width=30pc]{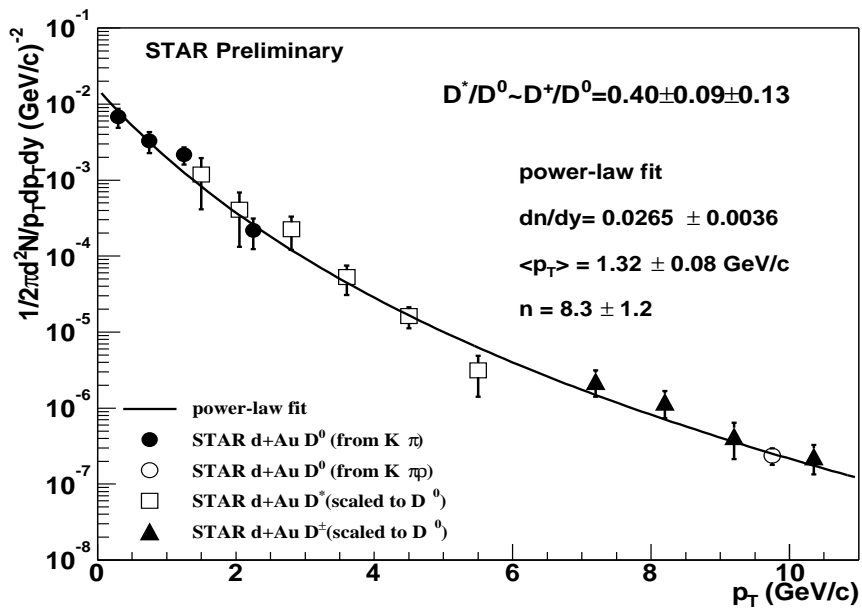}
\caption{The measured invariant yield distributions for $D^0$,
$D^{*\pm}$ and $D^\pm$ and a power-law fit to the data points. The
$D^*$ and $D^\pm$ data points were scaled by the ratio
$D^*/D^0=D^+/D^0=0.40\pm 0.09\pm 0.13$ after the power-law
fit.}\label{fig:ptspectra}
\end{figure}

The invariant yield distributions as a function of $p_T$ for
$D^{*\pm}$, $D^\pm$ and the high $p_T$ $D^0$ were obtained using
the same methods as that for the low $p_T$ $D^0$. With all the
data points from the direct open charm measurements shown in
Fig.~\ref{fig:ptspectra}, the $p_T$ distribution was fit to a
power-law function, $A(1+p_T/p_0)^{-n}$, with the ratio of
$D^*/D^0=D^+/D0$ as a free parameter, where $A$, $p_0$ and $n$ are
fit parameters. From the fit, we obtained
$dN/dy(D^0)$=0.027$\pm$0.004$\pm$0.007 which is consistent with
the exponential fit results to the low $p_T$ $D^0$ data points
only, $\langle p_T\rangle$=1.32$\pm$0.08$\pm$0.16 GeV/c, and the
ratio $D^*/D^0=D^+/D^0=0.40\pm 0.09\pm 0.13$ which agrees with the
ratios from previous measurements~\cite{e1} and theoretical
predictions~\cite{t1} within the experimental uncertainties.

\section{Conclusion}
The direct open charm $D^0$, $D^{*\pm}$ and $D^{\pm}$ signals were
reconstructed in d+Au collisions at $\sqrt{s_{NN}}$=200 GeV using
the STAR TPC. Single electron/positron spectra were measured using
the TOFr+TPC for $p_T<$3 GeV/c and using the TPC only for
2$<p_T<$4 GeV/c in d+Au and p+p collisions at the same beam
energy. From the low $p_T$ $D^0$ and the single electron
measurements, a total $c\bar{c}$ cross section was obtained and is
at least about a factor of 3 larger than PYTHIA and NLO pQCD
calculations. The $D^0$, $D^{*\pm}$ and $D^{\pm}$ open charm $p_T$
spectra in d+Au collisions was fit to a power-law fit and the
$D^*/D^0=D^+/D^0=0.40\pm 0.09\pm 0.13$ was extracted and agrees
with the ratios from previous measurements and theoretical
predictions within the experimental uncertainties.
\\

\end{document}